\font\cmbxti = cmbxti10	%
\font\cmmib  = cmmib10	%
\font\cmbxtitwelve = cmbxti10 scaled 1200
\font\cmmibtwelve = cmmib10 scaled 1200
\newfam\mibtenfam \textfont\mibtenfam=\cmmib \scriptfont\mibtenfam=\cmmib
	\scriptscriptfont\mibtenfam=\cmmib
\newfam\mibtwelvefam\textfont\mibtwelvefam=\cmmibtwelve
	\scriptfont\mibtwelvefam=\cmmibtwelve
	\scriptscriptfont\mibtwelvefam=\cmmibtwelve
\def\bfitten{\fam\mibtenfam\cmbxti}
\def\bfittwelve{\fam\mibtwelvefam\cmbxtitwelve}
\let\bfit\bfitten
\mathchardef\alpha="710B\mathchardef\beta="710C\mathchardef\gamma="710D
\mathchardef\delta="710E\mathchardef\epsilon="710F\mathchardef\zeta="7110
\mathchardef\eta="7111\mathchardef\theta="7112\mathchardef\iota="7113
\mathchardef\kappa="7114\mathchardef\lambda="7115\mathchardef\mu="7116
\mathchardef\nu="7117\mathchardef\xi="7118\mathchardef\pi="7119
\mathchardef\rho="711A\mathchardef\sigma="711B\mathchardef\tau="711C
\mathchardef\upsilon="711D\mathchardef\phi="711E\mathchardef\chi="711F
\mathchardef\psi="7120\mathchardef\omega="7121\mathchardef\varepsilon="7122
\mathchardef\vartheta="7123\mathchardef\varpi="7124\mathchardef\varrho="7125
\mathchardef\varsigma="7126\mathchardef\varphi="7127
\ifx\DAVIDRAMACROS\undefined\relax\else\endinput\fi
\let\DAVIDRAMACROS1
\catcode`@=11					%
\let\notglobal=\relax%
\let\notouter=\relax%
\notouter\def\newcount{\alloc@0\count\countdef\insc@unt}
\notouter\def\newdimen{\alloc@1\dimen\dimendef\insc@unt}
\notouter\def\newskip{\alloc@2\skip\skipdef\insc@unt}
\notouter\def\newmuskip{\alloc@3\muskip\muskipdef\@cclvi}
\notouter\def\newbox{\alloc@4\box\chardef\insc@unt}
\notouter\def\newtoks{\alloc@5\toks\toksdef\@cclvi}
\notouter\def\newhelp#1#2{\newtoks#1#1\expandafter{\csname#2\endcsname}}
\notouter\def\newread{\alloc@6\read\chardef\sixt@@n}
\notouter\def\newwrite{\alloc@7\write\chardef\sixt@@n}
\notouter\def\newfam{\alloc@8\fam\chardef\sixt@@n}
\notouter\def\newinsert#1{\global\advance\insc@unt by\m@ne
  \ch@ck0\insc@unt\count
  \ch@ck1\insc@unt\dimen
  \ch@ck2\insc@unt\skip
  \ch@ck4\insc@unt\box
  \allocationnumber=\insc@unt
  \global\chardef#1=\allocationnumber
  \wlog{\string#1=\string\insert\the\allocationnumber}}
\notouter\def\newif#1{\count@\escapechar \escapechar\m@ne
  \expandafter\expandafter\expandafter
   \edef\@if#1{true}{\let\noexpand#1=\noexpand\iftrue}%
  \expandafter\expandafter\expandafter
   \edef\@if#1{false}{\let\noexpand#1=\noexpand\iffalse}%
  \@if#1{false}\escapechar\count@} %
\def\alloc@#1#2#3#4#5{\global\advance\count1#1 by 1
  \ch@ck#1#4#2%
  \allocationnumber=\count1#1
  \notglobal#3#5=\allocationnumber
  \wlog{\string#5=\string#2\the\allocationnumber}}
\catcode`@=12					%
\newbox\IPtemp				%
\newdimen\IPtempd			%
\newdimen\IPspace\IPspace=.5em		%
\newdimen\IPleft\IPleft=0pt		%
\newif\ifIPtemps			%
\newif\ifnotIPtemps			%
\long\def\IP#1#2{{%
\setbox\IPtemp=\hbox{\noindent#1\hbox to \IPspace{}\hfil}%
\IPtempsfalse%
\notIPtempstrue%
\ifdim\IPleft>0pt%
	\ifdim\wd\IPtemp>\IPleft%
		\relax%
		\IPtempstrue%
		\notIPtempsfalse%
	\fi%
\fi%
\ifnotIPtemps\ifdim\wd\IPtemp<\IPleft%
	\setbox\IPtemp=\hbox{\hbox to \IPleft{\unhcopy\IPtemp}}%
\fi\fi%
\IPtempd=\the\hsize%
\ifnotIPtemps					%
	\advance\IPtempd -\wd\IPtemp		%
	\noindent%
	\hbox{\unhbox\IPtemp\vtop{\hsize=\IPtempd\noindent#2}}%
\else						%
	\advance\IPtempd -\IPleft%
	\noindent%
	\vbox{%
		\hbox{\unhbox\IPtemp}		%
		\hbox{\hbox to \IPleft{}\vtop{\hsize=\IPtempd\noindent#2}}%
	}%
\fi%
\par%
}}
\font\eightrm=cmr8
\def\bold#1{{\bf #1}}
\def\roman#1{{\rm #1}}
\parskip=\smallskipamount		%
\def\quote#1{{\vskip 12pt \moveright 40pt \vbox{\hsize=5truein
    \baselineskip=14pt \noindent #1}}}
\def\boxit#1{\vbox{\hrule\hbox{\vrule\kern3pt         %
     \vbox{\kern3pt#1\kern3pt}\kern3pt\vrule}\hrule}} %

\def\fr#1/#2{{\textstyle{#1\over#2}}} 
\def\frac#1/#2{\leavevmode\kern0em\raise.5ex\hbox{$\scriptstyle#1$}%
\kern-.15em/\kern-.15em\lower.25ex\hbox{$\scriptstyle#2$}}

\newcount\ftnumber
\long\def\ft#1{\ignorespaces\global\advance\ftnumber by 1
          {\baselineskip\normalbaselineskip
           \footnote{$^{\the\ftnumber}$}{\ignorespaces#1}}}
\long\def\nft#1{\ignorespaces\global\advance\ftnumber by 1
	\baselineskip\normalbaselineskip
	\footnote{{\sevenrm\raise0.8em\hbox{\the\ftnumber}}}{\ignorespaces#1}}
\def\ftname(#1){\expandafter\xdef\csname #1\endcsname{\the\ftnumber}}
\newlinechar=`\^^J%
\newcount\ranfoo%
\ranfoo=\day\multiply\ranfoo\deadcycles%
\advance\ranfoo\pagetotal\advance\ranfoo\time%
\edef\tempenfilnam{./tex\the\ranfoo.tex}%
\newwrite\tempenfile	%
\newwrite\tempenfile%
\newread\tempenrfile%
\newread\tempenrfile%
\newcount\ennumber\ennumber=0%
\long\def\endnote#1{%
	\newlinechar=`\^^J%
	\ifnum\number\ennumber=0\immediate\openout\tempenfile=\tempenfilnam\fi%
	\global\advance\ennumber1%
	\immediate\write\tempenfile{#1}%
}%
\newif\ifnotend%
\def\endnotes{%
	\def\xdef##1##2{\null}		%
	\catcode`@=11%
	\immediate\closeout\tempenfile%
	\immediate\openin\tempenrfile=\tempenfilnam%
	\loop%
		\ifeof\tempenrfile\notendfalse\else\notendtrue%
			\immediate\read\tempenrfile to \foobie%
			\foobie%
		\fi%
	\ifnotend\repeat%
	\catcode`@=11%
}%
\long\def\prlnote#1{%
	\endnote{\par$^{\the\ennumber}$ #1}%
	$^{\the\ennumber}$%
}%
\font\tiny = cmr5 scaled 500%
\newcount\eqnumber
\def\eq(#1){%
    \ifx\DRAFT\undefined\def\DRAFT{0}\fi	%
    \global\advance\eqnumber by 1%
    \expandafter\xdef\csname !#1\endcsname{\the\eqnumber}%
    \ifnum\number\DRAFT>0%
	\setbox0=\hbox{\tiny #1}%
	\wd0=0pt%
	\eqno({\offinterlineskip
	  \vtop{\hbox{\the\eqnumber}\vskip1.5pt\box0}})%
    \else%
	\eqno(\the\eqnumber)%
    \fi%
}
\def\silenteq(#1){%
    \global\advance\eqnumber by 1%
    \expandafter\xdef\csname !#1\endcsname{\the\eqnumber}%
}
\def\alteq(#1){%
    \eqno({\rm\the\eqnumber#1})%
}
\def\(#1){(\csname !#1\endcsname)}
\def\altq(#1@#2){({\rm\csname !#1\endcsname#2})}
\def\nexteqn(+#1){{%
\count0=\the\eqnumber%
\advance\count0 #1%
(\the\count0)%
}}
\newbox\FatInternalVariable		%
\def\fat#1{{
\kern-.30em%
\hbox{					%
\setbox\FatInternalVariable=\hbox{#1}%
\unskip%
\unhcopy\FatInternalVariable%
\kern-\wd\FatInternalVariable%
\kern 0.015 em%
\unhcopy\FatInternalVariable%
\kern-\wd\FatInternalVariable%
\kern 0.030 em%
\unhbox\FatInternalVariable%
}
\kern-.33em%
}}
\font\eightrm=cmr8 \font\eighti=cmmi8 \font\eightsy=cmsy8
\font\eightex=cmex10 scaled 800 %
\font\eightit=cmti8 \font\eightsl=cmsl8 \font\eighttt=cmtt8
\font\eightbf=cmbx8
\font\sixrm=cmr7 \font\sixi=cmmi6 \font\sixsy=cmsy6 \font\sixbf=cmbx6
\font\sixex=cmex10 scaled 600 %
\newskip\ttglue
\def\eightpoint{\def\rm{\fam0\eightrm}%
  \textfont0=\eightrm\scriptfont0=\sixrm\scriptscriptfont0=\fiverm%
  \textfont1=\eighti\scriptfont1=\sixi\scriptscriptfont1=\fivei%
  \textfont2=\eightsy\scriptfont2=\sixsy\scriptscriptfont2=\fivesy%
  \textfont3=\eightex\scriptfont3=\sixex\scriptscriptfont3=\sixex%
  \textfont\itfam=\eightit \def\it{\fam\itfam\eightit}%
  \textfont\slfam=\eightsl \def\sl{\fam\slfam\eightsl}%
  \textfont\ttfam=\eighttt \def\tt{\fam\ttfam\eighttt}%
  \textfont\bffam=\eightbf \def\bf{\fam\bffam\eightbf}%
  \scriptfont\bffam=\sixbf%
  \scriptscriptfont\bffam=\fivebf%
  \tt\ttglue=.5em plus.25em minus.15em%
  \normalbaselineskip=9pt%
  \setbox\strutbox=\hbox{\vrule height7pt depth2pt width0pt}%
  \ifx\bfitten\undefined\relax\else\let\bfit\bfitten\fi%
  \let\sc=\sixrm\let\big=\eightbig\normalbaselines\rm}%
\font\twelverm=cmr12 \font\twelvei=cmmi12 \font\twelvesy=cmsy10 scaled 1200
\font\twelveex=cmex10 scaled 1200 %
\font\twelveit=cmti12 \font\twelvesl=cmsl12 \font\twelvett=cmtt12
\font\twelvebf=cmbx12
\font\twrmsc=cmr10 scaled960%
\font\twrmscsc=cmr8 scaled900%
\font\twisc=cmmi10 scaled960%
\font\twiscsc=cmmi8 scaled900%
\font\twsysc=cmsy10 scaled960%
\font\twsyscsc=cmsy8 scaled900%
\font\twexsc=cmex10 scaled960%
\font\twexscsc=cmex8 scaled900%
\def\twelvepoint{\def\rm{\fam0\twelverm}%
  \textfont0=\twelverm\scriptfont0=\twrmsc\scriptscriptfont0=\twrmscsc%
  \textfont1=\twelvei\scriptfont1=\twisc\scriptscriptfont1=\twiscsc%
  \textfont2=\twelvesy\scriptfont2=\twsysc\scriptscriptfont2=\twsyscsc%
  \textfont3=\twelveex\scriptfont3=\twexsc\scriptscriptfont3=\twexscsc%
  \textfont\itfam=\twelveit \def\it{\fam\itfam\twelveit}%
  \textfont\slfam=\twelvesl \def\sl{\fam\slfam\twelvesl}%
  \textfont\ttfam=\twelvett \def\tt{\fam\ttfam\twelvett}%
  \textfont\bffam=\twelvebf \def\bf{\fam\bffam\twelvebf}%
  \scriptfont\bffam=\tenbf%
  \scriptscriptfont\bffam=\eightbf%
  \tt\ttglue=.75em plus.37em minus.22em%
  \normalbaselineskip=14pt%
  \setbox\strutbox=\hbox{\vrule height10.5pt depth3pt width0pt}%
  \ifx\bfittwelve\undefined\relax\else\let\bfit\bfittwelve\fi%
  \let\sc=\sixrm\let\big=\twelvebig\normalbaselines\rm}
\def\tenpoint{\def\rm{\fam0\tenrm}%
  \textfont0=\tenrm\scriptfont0=\eightrm\scriptscriptfont0=\sixrm%
  \textfont1=\teni\scriptfont1=\eighti\scriptscriptfont1=\sixi%
  \textfont2=\tensy\scriptfont2=\eightsy\scriptscriptfont2=\sixsy%
  \textfont3=\tenex\scriptfont3=\eightex\scriptscriptfont3=\sixex%
  \textfont\itfam=\tenit \def\it{\fam\itfam\tenit}%
  \textfont\slfam=\tensl \def\sl{\fam\slfam\tensl}%
  \textfont\ttfam=\tentt \def\tt{\fam\ttfam\tentt}%
  \textfont\bffam=\tenbf \def\bf{\fam\bffam\tenbf}%
  \scriptfont\bffam=\eightbf%
  \scriptscriptfont\bffam=\sixbf%
  \tt\ttglue=.625em plus.31em minus.185em%
  \normalbaselineskip=11.5pt%
  \setbox\strutbox=\hbox{\vrule height8.75pt depth2.5pt width0pt}%
  \ifx\bfitten\undefined\relax\else\let\bfit\bfitten\fi%
  \let\sc=\sixrm\let\big=\tenbig\normalbaselines\rm}
\font\elevenrm=cmr10 scaled 1100\font\eleveni=cmmi10 scaled 1100%
\font\elevensy=cmsy10 scaled 1100
\font\elevenex=cmex10 scaled 1100 %
\font\elevenit=cmti10 scaled 1100
\font\elevensl=cmsl10 scaled 1100
\font\eleventt=cmtt10 scaled 1100
\font\elevenbf=cmbx10 scaled 1100
\font\ninerm=cmr9 %
\font\ninebf=cmbx9
\font\ninei=cmmi9\font\ninesy=cmsy9\font\nineex=cmex9
\def\elevenpoint{\def\rm{\fam0\elevenrm}%
  \textfont0=\elevenrm\scriptfont0=\ninerm\scriptscriptfont0=\eightrm%
  \textfont1=\eleveni\scriptfont1=\ninei\scriptscriptfont1=\eighti%
  \textfont2=\elevensy\scriptfont2=\ninesy\scriptscriptfont2=\eightsy%
  \textfont3=\elevenex\scriptfont3=\nineex\scriptscriptfont3=\eightex%
  \textfont\itfam=\elevenit \def\it{\fam\itfam\elevenit}%
  \textfont\slfam=\elevensl \def\sl{\fam\slfam\elevensl}%
  \textfont\ttfam=\eleventt \def\tt{\fam\ttfam\eleventt}%
  \textfont\bffam=\elevenbf \def\bf{\fam\bffam\elevenbf}%
  \scriptfont\bffam=\ninebf%
  \scriptscriptfont\bffam=\eightbf%
  \tt\ttglue=.6875em plus.34em minus.2025em%
  \normalbaselineskip=12.75pt%
  \setbox\strutbox=\hbox{\vrule height9.625pt depth2.75pt width0pt}%
  \ifx\bfiteleven\undefined\relax\else\let\bfit\bfiteleven\fi%
  \let\sc=\sixrm\let\big=\elevenbig\normalbaselines\rm}
\def\bold#1{{\bf#1}}

\def\lheader#1{\bigskip\leftline{\bf#1}\nobreak}
\def\header#1{\bigskip\centerline{\bf#1}\nobreak}
\def\subsection#1#2{\header{#1. #2}}

\long\def\defbox#1=#2{\newbox#1\setbox#1=#2} %
\def\defdimen#1=#2{\newdimen#1#1=#2}

\long\def\VV#1#2{{\hbox{%
\newbox\VVA\newbox\VVB%
\setbox\VVA=\hbox{#2}%
\newdimen\VVC\VVC=\wd\VVA%
\setbox\VVB=\hbox{\vbox to#1{\vfil\hsize=\VVC\hbox{\unhbox\VVA}\vfil}}%
\wd\VVB=\VVC%
\divide\VVC2%
\unhbox\VVB}}}

\def\={\equiv}

\def\seq(#1){\eq(\SEC.#1)}
\def\seqr(#1){\(\SEC.#1)}
\defdimen\fakeIPleft=0pt%
\def\fakeIPinsert{\relax}%
\long\def\fakeIP#1#2{{\defdimen\oldleft=\leftskip%
\defdimen\oldparindent=\the\parindent%
\long\def\fakeft##1{{\leftskip\oldleft\parindent=\the\oldparindent\ft{##1}}}%
\defbox\bletch=\hbox{\noindent#1\hbox to \IPspace{}\hfil}%
\ifdim\fakeIPleft<\wd\bletch\relax\fakeIPleft=\wd\bletch\relax\fi%
\global\leftskip=\fakeIPleft\noindent%
\llap{\hbox to\leftskip{#1\hfil}}%
\fakeIPinsert%
\noindent\parindent=0pt%
\parindent=0pt%
\ignorespaces%
\def\fakeIPkludge{%
\displayindent=\leftskip\advance\displaywidth-\displayindent}%
{\noindent\ignorespaces#2}\par\global\leftskip=\oldleft}} %
\long\def\cstok#1{\def\hsp{\hphantom{.}}%
	\def\vsp{\kern2pt}%
	\leavevmode\hsp\hbox{\vrule\vtop{\vbox{\hrule\kern1pt\vsp%
	\hbox{\vphantom{/}\hsp{#1}\hsp}}%
	\kern1pt\vsp\hrule}\vrule}\hsp}

\newread\epsffilein    %
\newif\ifepsffileok    %
\newif\ifepsfbbfound   %
\newif\ifepsfverbose   %
\newif\ifepsfdraft     %
\newdimen\epsfxsize    %
\newdimen\epsfysize    %
\newdimen\epsftsize    %
\newdimen\epsfrsize    %
\newdimen\epsftmp      %
\newdimen\pspoints     %
\pspoints=1bp          %
\epsfxsize=0pt         %
\epsfysize=0pt         %
\def\epsfbox#1{\global\def\epsfllx{72}\global\def\epsflly{72}%
   \global\def\epsfurx{540}\global\def\epsfury{720}%
   \def\lbracket{[}\def\testit{#1}\ifx\testit\lbracket
   \let\next=\epsfgetlitbb\else\let\next=\epsfnormal\fi\next{#1}}%
\def\epsfgetlitbb#1#2 #3 #4 #5]#6{\epsfgrab #2 #3 #4 #5 .\\%
   \epsfsetgraph{#6}}%
\def\epsfnormal#1{\epsfgetbb{#1}\epsfsetgraph{#1}}%
\def\epsfgetbb#1{%
\openin\epsffilein=#1
\ifeof\epsffilein\errmessage{I couldn't open #1, will ignore it}\else
   {\epsffileoktrue \chardef\other=12
    \def\do##1{\catcode`##1=\other}\dospecials \catcode`\ =10
    \loop
       \read\epsffilein to \epsffileline
       \ifeof\epsffilein\epsffileokfalse\else
          \expandafter\epsfaux\epsffileline:. \\%
       \fi
   \ifepsffileok\repeat
   \ifepsfbbfound\else
    \ifepsfverbose\message{No bounding box comment in #1; using defaults}\fi\fi
   }\closein\epsffilein\fi}%
\def\epsfclipoff{\def\epsfclipstring{\ifepsfdraft\space clip\fi}}%
\epsfclipoff
\def\epsfsetgraph#1{%
   \epsfrsize=\epsfury\pspoints
   \advance\epsfrsize by-\epsflly\pspoints
   \epsftsize=\epsfurx\pspoints
   \advance\epsftsize by-\epsfllx\pspoints
   \epsfxsize\epsfsize\epsftsize\epsfrsize
   \ifnum\epsfxsize=0 \ifnum\epsfysize=0
      \epsfxsize=\epsftsize \epsfysize=\epsfrsize
      \epsfrsize=0pt
     \else\epsftmp=\epsftsize \divide\epsftmp\epsfrsize
       \epsfxsize=\epsfysize \multiply\epsfxsize\epsftmp
       \multiply\epsftmp\epsfrsize \advance\epsftsize-\epsftmp
       \epsftmp=\epsfysize
       \loop \advance\epsftsize\epsftsize \divide\epsftmp 2
       \ifnum\epsftmp>0
          \ifnum\epsftsize<\epsfrsize\else
             \advance\epsftsize-\epsfrsize \advance\epsfxsize\epsftmp \fi
       \repeat
       \epsfrsize=0pt
     \fi
   \else \ifnum\epsfysize=0
     \epsftmp=\epsfrsize \divide\epsftmp\epsftsize
     \epsfysize=\epsfxsize \multiply\epsfysize\epsftmp   
     \multiply\epsftmp\epsftsize \advance\epsfrsize-\epsftmp
     \epsftmp=\epsfxsize
     \loop \advance\epsfrsize\epsfrsize \divide\epsftmp 2
     \ifnum\epsftmp>0
        \ifnum\epsfrsize<\epsftsize\else
           \advance\epsfrsize-\epsftsize \advance\epsfysize\epsftmp \fi
     \repeat
     \epsfrsize=0pt
    \else
     \epsfrsize=\epsfysize
    \fi
   \fi
   \ifepsfverbose\message{#1: width=\the\epsfxsize, height=\the\epsfysize}\fi
   \epsftmp=10\epsfxsize \divide\epsftmp\pspoints
   \vbox to\epsfysize{\vfil\hbox to\epsfxsize{%
      \ifnum\epsfrsize=0\relax
        \includegraphics{\ifepsfdraft}%
      \else
        \epsfrsize=10\epsfysize \divide\epsfrsize\pspoints
        \includegraphics{\ifepsfdraft}%
      \fi
      \hfil}}%
\global\epsfxsize=0pt\global\epsfysize=0pt}%
{\catcode`\%=12 \global\let\epsfpercent=
\long\def\epsfaux#1#2:#3\\{\ifx#1\epsfpercent
   \def\testit{#2}\ifx\testit\epsfbblit
      \epsfgrab #3 . . . \\%
      \epsffileokfalse
      \global\epsfbbfoundtrue
   \fi\else\ifx#1\par\else\epsffileokfalse\fi\fi}%
\def\epsfempty{}%
\def\epsfgrab #1 #2 #3 #4 #5\\{%
\global\def\epsfllx{#1}\ifx\epsfllx\epsfempty
      \epsfgrab #2 #3 #4 #5 .\\\else
   \global\def\epsflly{#2}%
   \global\def\epsfurx{#3}\global\def\epsfury{#4}\fi}%
\def\epsfsize#1#2{\epsfxsize}

\def\fft#1{\ft{#1}\endnote{\par[\the\ennumber]\ #1}}
\def\figspace{1}		%
\def\figclust{2}		%
\def\figchij{3}			%
\def\figfit{4}			%
\def\figchis{5}			%
\def\figentropic{6}		%
\def\figfoursite{7}		%
\long\def\efigureinsert#1#2#3#4{{%
\midinsert%
\centerline{#3\epsfbox{#1}}%
\vskip\baselineskip%
\gdef#2{#4}%
\centerline{$\vcenter{\vbox{\hsize0.85\hsize\tenpoint\baselineskip1.19em%
\noindent#2}}$}%
\endinsert%
}}

\long\def\tableinsert#1#2#3#4{{%
\gdef#1{#4}%
\gdef#2{#3}%
\let\knuthisanidiot##\relax%
\midinsert%
\centerline{$\vcenter{\vbox{#3}}$}%
\vskip\baselineskip%
\centerline{$\vcenter{\vbox{\hsize0.85\hsize\tenpoint\baselineskip1.19em%
\noindent#4}}$}%
\endinsert%
}}
\long\def\tablenoinsert#1#2#3#4{{%
\gdef#1{#4}%
\gdef#2{#3}%
\let\knuthisanidiot##\relax%
\centerline{$\vcenter{\vbox{#3}}$}%
\vskip\baselineskip%
\centerline{$\vcenter{\vbox{\hsize0.85\hsize\tenpoint\baselineskip1.19em%
\noindent#4}}$}%
}}
\font\msxm msam10\relax%
\textfont"E\msxm%
\mathchardef\geo"3E26%
\mathchardef\leo"3E2E%
\def\kbt{k_{\hbox{\eightpoint B}}T}
\twelvepoint
\baselineskip1.5\baselineskip
\ %
\vskip -0.35cm
\nopagenumbers		%
\font\title cmbx12 scaled 1200
\def\small#1{{\twelverm #1}\tenpoint}
\centerline{\title A Spin Model for Investigating Chirality}
\vskip 0.18cm
\centerline{\small BY \small DAVID \small A. \small RABSON
AND \small STUART \small A. \small TRUGMAN}
\centerline{\sl MS B-262, Condensed-Matter Theory Group, T-11}
\centerline{\sl Los Alamos National Laboratory, Los Alamos, NM 87545, USA}
\vskip 0.52cm

{
\baselineskip.75\baselineskip

\centerline{\bf Abstract}\nobreak\vskip.2\baselineskip\nobreak
\hbox to\hsize{\hfil\vbox{\hsize=0.76\hsize\noindent%
Spin chirality has generated great interest recently both from
possible applications to flux phases
and intrinsically, as an example of a several-site
magnetic order parameter that can be long-ranged even
where simpler order parameters are not.  Previous work (motivated
by the flux phases) has focused on antiferromagnetic chiral order; we
construct a model in which the chirality orders ferromagnetically and
investigate the model's behavior as a function of spin.  Enlisting the aid of
exact diagonalization, spin-waves, perturbation theory, and mean fields,
we conclude that the model likely has long-ranged chiral order for spins
$1$ and greater and no non-trivial chiral order for spin \frac1/2.
}\hfil}

\vskip -0.1cm\nobreak\noindent

}

\vskip 0.17cm
\centerline{May 1995}
\vskip 0.17cm
\centerline{\bf submitted to {\bfit Journal of Physics: Condensed Matter}}
\vskip 0.0cm
\noindent
{\tt PACS-95 75.10.Jm, 75.30.Ds}
\vskip 1cm
\noindent{Los Alamos Unclassified Release (LAUR) number 95-1734}
\vfil
\eject
\pageno=1
\footline={\hss\tenrm\folio\hss}

\pageno1\relax				%
\footline={\hss\twelverm\folio\hss}	%
\header{Chirality, a Simple Model, and its Classical Ground State}
Among the flood of theories that has issued from the discovery
eight years ago of a new class of superconductors, several
concern themselves with a competition between antiferromagnetic
and chiral order.\fft{%
See {\it e.g.} Wen, Wilczek, and Zee, 1989; Baskaran, 1989; Gooding, 1990;
Marston and Zeng, 1991; Korshunov, 1993; Masui and Betts, 1993; 
and references contained
therein.\ftname(footchiralrefs)} %
These efforts have focused on antiferromagnetic Hamiltonians, often
involving frustration.  We examine a different way of arriving at
chiral order, one using Heisenberg-squared bonds
$ ( \bold S_i \cdot \bold S_j )^2 $
(which, for
spin greater than \frac1/2, cannot be reduced to Heisenberg
terms).\fft{%
We note an interest in this ``biquadratic'' exchange (so
called for its second-quantized expression) independent of any
resultant chirality: Anderson (1963) (page 168);
Harris and Owen (1963); Takhtajan (1982);
Affleck {\it et al.} (1987); Mitra and Chakraborty (1994).} %

Given spins $i$, $j$, and $k$
forming the vertices of a triangle
(in figure \figspace, half a square of a square lattice), we
define the (pseudoscalar) chiral order parameter as
$$
\chi_{ijk} = \bold S_i \cdot ( \bold S_j \times \bold S_k )\smash{\quad.}
\eq(chidef)
$$

\efigureinsert{figspace.ips}{\capspace}{\epsfysize=3in}{%
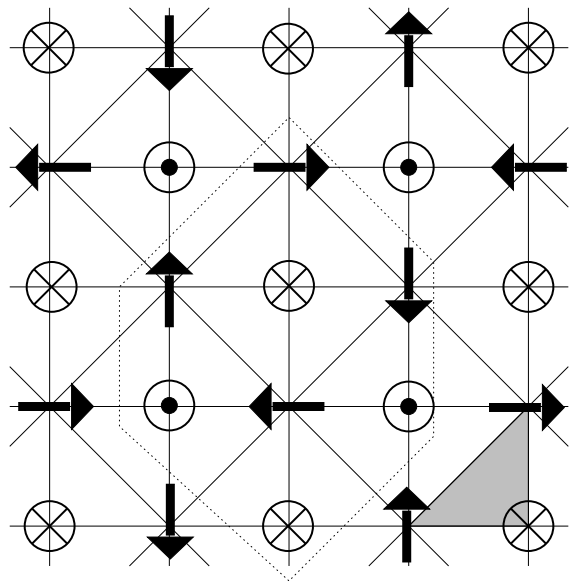
{\bf Figure \figspace.}  The geometry of the problem: right-, left-, up-,
down-, outward- ($\odot$), and inward-pointing ($\otimes$)
arrows give one classical ground state
(others are related by global spin rotation and chirality flip).  The
lines indicate Heisenberg-squared $(\bold S_i\cdot\bold S_j)^2$ bonds;
the antiferromagnetic bonds between opposite-pointing spins are not shown.
The sublattice pictured here with spins pointing into or out of the page
we call ``even'' or sublattice $0$; sublattices $1$, with spins pointing
left or right, and $2$, up and down, we call ``odd.''
The dotted lines highlight a ground-state unit cell.  One example of the
plaquettes on which we define chirality
is shaded in the lower-right corner.
}
We can consider the spin $\bold S$
either a quantum spin-$s$ or a classical (vector) object; the latter
is equivalent to the limit $s\rightarrow\infty$.
We will begin with classical
spins, constructing interactions %
in such manner as to
ensure a long-range-ordered chiral ground state, then investigate
the quantum behavior of the resulting model as we lower the
spin from $\infty$ to $1/2$.  Understanding the ground state
as a function of spin may later prove useful in more realistic
models.\fft{%
As an example, $\roman U\,\roman R\roman u_2\,\roman S\roman i_2$ has a large
specific-heat jump at a presumably magnetic transition,
yet almost no antiferromagnetic
ordering, a discrepancy
Gor'kov and Sokol (1992) and Agterberg and Walker (1994) explain
with chiral order parameters.} %

For convenience, we
use a square lattice.  There are two translationally-related ways
of drawing diagonals
across squares to split the lattice into triangular plaquettes
while preserving the point group of the whole lattice.
We will construct a Hamiltonian such that $\chi$ on the triangles
outlined in figure \figspace\ takes its maximal absolute value, while on
the triangles formed by drawing diagonals the other way it has zero
value.  We note that this will result in chiral-ferromagnetic order,
in that $\chi$ has the same sign everywhere on the lattice, as opposed
to the chiral-antiferromagnetism considered in the references listed
in footnote \footchiralrefs.  (Antiferromagnetic chirality is motivated
in part by an analogy to the flux phase of Marston and Affleck, 1989, in which 
a gauge field living on the bonds of a square lattice induces a staggered
magnetic-like field on the square plaquettes.  A ferromagnetic chirality,
in contrast, would correspond to a gauge field that cancels everywhere in
the interior, leaving a net circulation around the outside of a sample.)

Viewed as classical vectors, the spins on the vertices of a triangle
maximize the absolute value of $\chi$ for that triangular plaquette
when they are mutually perpendicular: we therefore begin with a
Heisenberg-squared term
$(\bold S_i \cdot \bold S_j)^2$
for all neighboring sites $i, j$ and those pairs of next-neighbor sites joined
by diagonals in figure \figspace.  By itself, however, this would yield local
chiral plaquettes with no correlation in sign between one plaquette and
its neighbors.  To enforce such correlation, it suffices to add
antiferromagnetic bonds linking those nearest spins that we wish (see figure
\figspace) to align antiferromagnetically: specifically, we add
next-neighbor antiferromagnetic bonds across the square diagonals that do
not already have Heisenberg-squared bonds and between ``odd'' third-neighbor
pairs, where we define as odd those sites in figure \figspace\ from
which emanate
eight rays against the four at each even site.  To maintain the two
kinds of terms, those quartic in spin and those quadratic,
equal in strength for all values of spin, we divide each term by
spin to the appropriate power:
$$
H=J s^{-4}\sum_{\eightpoint\hbox to2em{\hss${\hbox{Heisenberg}^2\ i,j}$\hss}}
(\bold S_i\cdot\bold S_j)^2
+s^{-2}\sum_{\eightpoint\hbox to2em{\hss${\hbox{anti-ferro.\ }%
^{\vphantom{2}}i,j}$\hss}}
\bold S_i\cdot\bold S_j\smash{\quad.}
\eq(Hamiltonian)
$$
With this Hamiltonian,
one easily confirms the classical
ground state of figure \figspace, up to a global spin rotation and the
global sign of chirality (plus or minus).

The parameter $J$ controls the relative importance of the quartic and
quadratic terms.  For classical spins, the ground state is the same for
any positive value.  With the Heisenberg-squared bonds turned off,
$J\rightarrow0$, the problem splits into three entirely independent
antiferromagnetic square sublattices.

The chirality, \(chidef), is a pseudoscalar, and the Hamiltonian locally
encourages it to take its full value, either positive or negative.  In
this sense, it more closely resembles an Ising than a vector spin:
even as long-wavelength twisting modes might cause spin-spin correlations
within each antiferromagnetic sublattice to
decay exponentially with distance, the
individual sublattices could remain locally mutually perpendicular, preserving
the sign and strength of the chirality to a higher temperature.
We therefore expect
that chiral order will be more robust than spin order unless
quantum-mechanical peculiarities at small spin favor another phase.

At zero temperature, the antiferromagnetic bonds
alone ($J=0$) suffice to order the ground state in a trivial way: with
the three decoupled sublattices (left-right, up-down, in-out
in figure \figspace) free to assume any relative orientation,
long-range antiferromagnetic order on each will correlate $\chi$ at
one plaquette with $\chi$ on any plaquette arbitrarily far away.
As we shall demonstrate, however, thermal fluctuations destroy this
freedom (``order from disorder''\ft{%
Villain, 1979,
Villain {\it et al.} 1980, Henley 1989.}) and with it, for spin-\frac1/2, the
chirality.
For spins 1 and
higher, the same fluctuations instead {\it favor\/} chiral order.  In the
zero-temperature diagonalizations to follow, we shall search for
evidence of ordering 1.~{\it in excess of the trivial order} and
2.~{\it for $J\geo1$, away from where we expect the trivial mechanism
to operate}.

\header{Exact Diagonalization of Small Clusters}
The classical ($s=\infty$) ground state is fixed, by our construction.
We examine in three ways the question of what happens as $s$ is lowered to more
physical values: first by exactly diagonalizing the problem for small
numbers of sites, then in a spin-wave (``$1/s$'') calculation, and finally
by considering related problems motivated by lingering doubts on
the fate of spin one-half.

We impose periodic boundary conditions.
While there are only two inequivalent sites in the Hamiltonian,
the classical ground state (see figure \figspace) has a unit cell of eight;
we would expect (and indeed verify) that any classically frustrated
cluster (for instance, one with other than a multiple of eight sites)
would have depressed chiral order.
Because the size
of the Hilbert space grows exponentially with the number of sites
and very fast with the spin as well, we cannot examine with
exact diagonalization clusters as
large we might wish.  Without loss of generality, we limit our
diagonalization to the space with zero total spin projected along
the ${\bf\hat z}$ axis
($ S_{\hbox{\eightpoint\rm total}}^z
= \sum_{\hbox{\eightpoint\rm sites\ }i} S_i^z = 0 $).
We furthermore assume the non-degeneracy of the ground state, so that
we may limit our attention to momenta 0 and $\pi$ along the two
spatial directions.\fft{%
This assumption holds for clusters small enough that
we could vary momentum.
} %
While in principle we could further reduce the size of the space
by restricting ourselves to the {\it total\/} spin-zero sector
($ \bold S_{\hbox{\eightpoint\rm total}} = \bold 0 $\nobreak),
employing the valence-bond (singlet) basis (Ramasesha and Soos, 1984;
Chang {\it et al.}, 1989), its
non-orthogonality makes it impractical for calculating
eigenvectors or correlation functions.\fft{%
The aforementioned valence-bond basis works well for some
one-dimensional problems with local (nearest or at worst second- or
third-neighbor)
interactions.  However, to evaluate two-dimensional interactions requires
either an expansion of the operators (see Ramasesha and Soos) or
the application of a diagram-untangling rule.  The latter is more
efficient but still results in the generation of a large number of
diagrams that must be contracted.  More significantly, the non-orthogonality
of the valence-bond basis converts the eigenproblem into the diagonalization
of a {\it non-Hermitian\/} matrix, precluding the use of the Lanczos
algorithm.  (Existing non-symmetric Lanczos algorithms do not work well,
while orthogonalizing the basis would prove as expensive as diagonalizing
the Hamiltonian, as well as destroying sparseness.)
Once the ground-state wavefunction has been found, nonorthogonality makes the
evaluation of observable operators an order-$n^2$ operation ($n$ the basis
size), destroying the advantage of shrinking the Hilbert space.
The difficulties become worse with larger spin: the
generalized basis described by Chang {\it et al.} (1989), while
appropriate for calculating energies, is unsuitable for
the evaluation of expectation values and must first be expanded.  The
number of expansion steps grows exponentially with the number of sites
and so is not practical.
} %
Once a basis with the proper symmetry has been constructed (following
the usual procedure of summing with proper phases the states in the
symmetry element's orbit), it is straightforward to find the ground state
with the Lanczos method.

\efigureinsert{figclust.ips}{\capclust}{\epsfysize=3in}{%
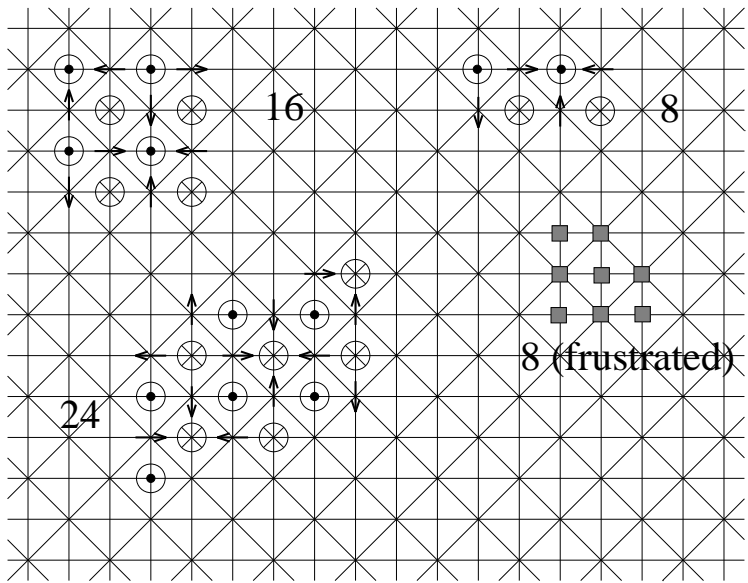
{\bf Figure \figclust.}  Exact diagonalization: we use clusters of
8, 16, and 24 spins, for which we show classical ground states.  For
purposes of comparison, we also use an additional eight-site cluster
frustrated by periodic boundary conditions.
}

For unfrustrated 8-, 16-, and 24-site spin-\frac1/2 clusters
(figure \figclust),
and for a range of spins and ratios $J$ of the two bond strengths,
we calculated the chirality-chirality correlations at the largest
separations possible, with the results shown in
figure \figchij.  The spin-\frac1/2 clusters (two bottom traces)
show their largest correlation, still less than 15\% of the
maximum attainable by three spins one-half, for negative $J$.
This value lies barely above the $J=0$ correlation of 0.1361 (16 spins)
of entirely uncoupled spin-\frac1/2 sublattices.  Since
the actual peak ordering occurs for $J\!<\!0$ and does not rise substantially
above the ``trivial'' ordering value, we conclude that spin-\frac1/2
has no non-trivial chiral ordering.\ft{%
For uncorrelated but ordered classical sublattices, averaging over relative
angles gives $\sqrt{\chi\chi}=\sqrt{2/9}\approx47\%$.
Quantum fluctuations reduce this to .1113 (normalized) in spin-wave theory for
spins \frac1/2, close to the .1361 quoted above and calculated for the
actual antiferromagnetic correlations measured in the cluster.  We
shall later provide evidence that in the small-$J$ regime that has
the ``trivial'' order, thermal fluctuations will destroy it.
\ftname(sqa)%
} %

\efigureinsert{figchij.ps4}{\capchij}{\epsfysize=3.6in}{%
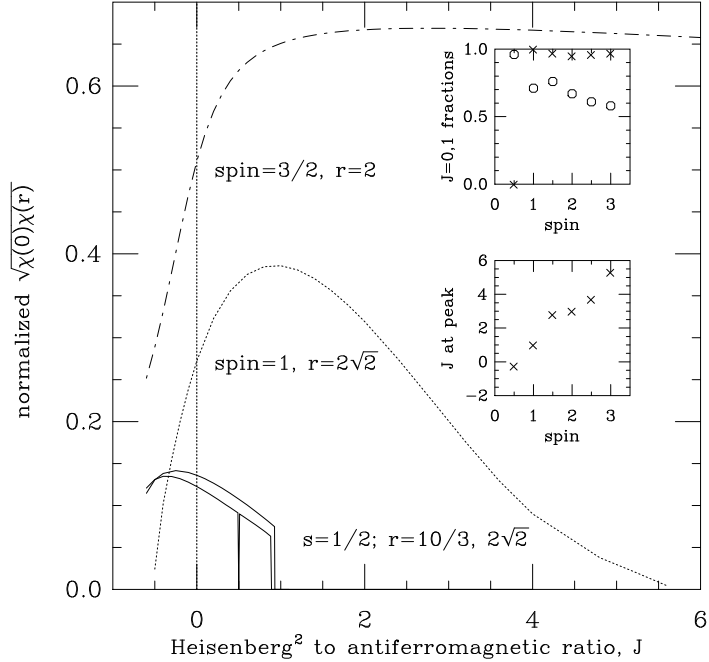
{\bf Figure \figchij.}  Exact diagonalizations varying the ratio $J$ of
the two bond strengths for spins
\frac1/2, 1, and \frac3/2: we
plot the square root of normalized
chirality-chirality correlation as a function of 
$J$.  Unity on the vertical axis represents the maximum eigenvalue of
$\chi(123)$.
The two lowest plots (solid lines), showing less than
15\% maximum chirality, are for
spin \frac1/2.  Note that the maximum occurs below $J=0$:
in other words, without the benefit of Heisenberg-squared coupling.  This level
of chirality is expected for uncorrelated antiferromagnetic lattices
and should be subtracted (see footnote \sqa).
The lower of the two solid traces is for 24 spins (it dips
once at $J=.5$, where a ferromagnetic state is exactly degenerate) and
a distance of 10/3, the largest possible in the cluster.  The
upper spin-\frac1/2 trace is for 16 sites and a distance of $2\sqrt{2}$.
These curves drop suddenly to zero correlation at $J=.89$ (24 sites) or
$.93$ (16 sites), corresponding to a partial ferromagnetic transition.
The ground state becomes fully ferromagnetic around $1.0$.
The next trace up (dotted line) is for 16 spin-1 sites; it peaks around
$J=1$, where it shows nearly half again as much chiral order as when zero $J$
decouples the sublattices.  Note that the order falls continuously to zero as
$J$ approaches about 6; there is no ferromagnetic transition for $J$ as
large as $100$.
We take this curve as evidence of at least
short-range order, possibly long-ranged.  The top curve (dashed) shows
the chirality correlation at a distance of $2$ for spin-\frac3/2 on
an eight-site cluster.

The $J$ corresponding to the peak correlation shifts higher
with larger spin, as indicated in the lower inset for spins up to
3.  Countering this trend, however, the curves become increasingly flat.
As shown by the crosses ($\times$) in the upper inset,
for spins greater than \frac1/2, the correlation at $J\!=\!1$ appears
to saturate near to the maximum for each spin.  Conversely, the
fraction of peak correlation realized with the Heisenberg-squared coupling
turned off ($J\!=\!0$, {\tenpoint$\bigcirc$}) starts
close to unity for spin-\frac1/2 but quickly
drops to a level compatible with random orientations.  Since the classical
problem ($\hbox{spin}\rightarrow\infty$) would have full correlation for
all $J\!>\!0$ but only 47\% correlation for $J\!=\!0$ (see footnote \sqa),
we take this inset as evidence
for a possible order-disorder transition between spin-1 and spin-\frac1/2.
}%

As $J$ increases,
the correlations fall, dropping abruptly to zero at $.93$ (for 16 spins,
or $.89$ for 24), coincident with a partial ferromagnetic transition.
While for the non-zero part of the curve the ground state forms a
spin singlet, in an intermediate regime ($J=$ about $.93$--$1$ for 16 spins,
$.89$--$1$ for 24), the cluster carries spin
one less than ferromagnetic,
spin $7$ for 16 sites or $11$ for 24.  Above $J=1$, both clusters become fully
ferromagnetic.
It is not surprising
that the Hamiltonian should behave differently for spin \frac1/2 than
classically, for
the Heisenberg-squared term $(\bold S_i\cdot\bold S_j)^2$ reduces
in the former case
to a simple ferromagnetic interaction,
${3\over16}-{1\over2}\bold S_i\cdot \bold S_j$.
Even at the value of $J$ for which the correlation reaches
a maximum, extrapolation of the correlation as a function
of distance (figure \figfit) is compatible with
no more than the trivial chirality
correlations expected for essentially decoupled lattices.

\efigureinsert{figfit.ps4}{\capfit}{\epsfysize=3.6in}{%
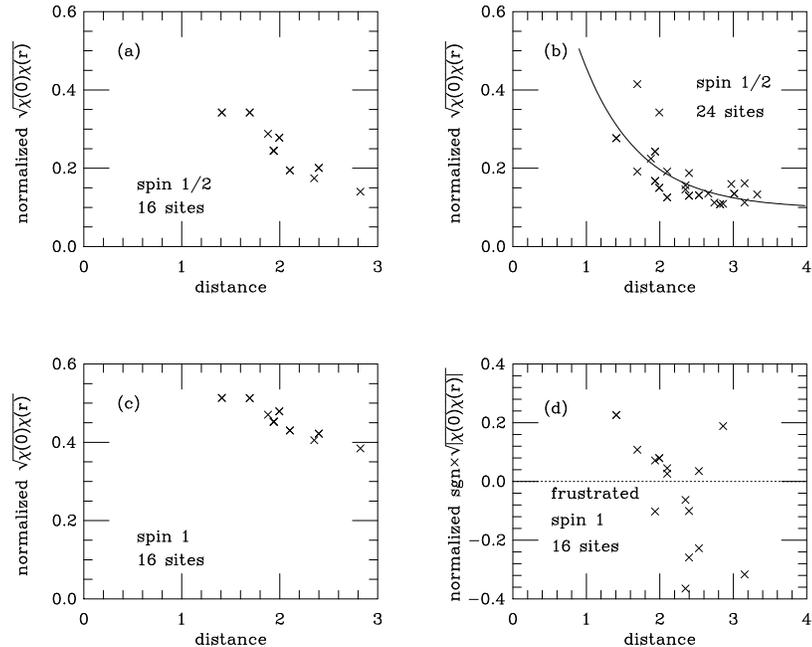
{\bf Figure \figfit.}  Distance dependence of chirality correlations for
spins \frac1/2 (upper row) and 1 (lower).
The vertical axes are normalized as in figure \figchij.
Least-squares fits to $c_0 + \gamma\exp(-x/\xi)$, $x$ the distance,
were inconclusive for the 16-site clusters.
(The root-mean-square variances were larger than the fitting parameters.)
However, from (b) we see 24 already approaching
an asymptotic correlation $c_0=0.10\pm.04$,
compatible with the predicted
$0.1113$ (footnote \sqa), and rapidly, with a coherence length
$\xi=.8\pm.4$ equal to or smaller than a lattice spacing.
Each case (a)--(c)
fixes $J$ at its peak value from figure \figchij.  In (d) we
plot the correlation for a frustrated spin-1 cluster.  Note the smaller
absolute correlations as well as the greater scatter.
}%

\efigureinsert{figchis.ps4}{\capchis}{\epsfysize=3.6in}{%
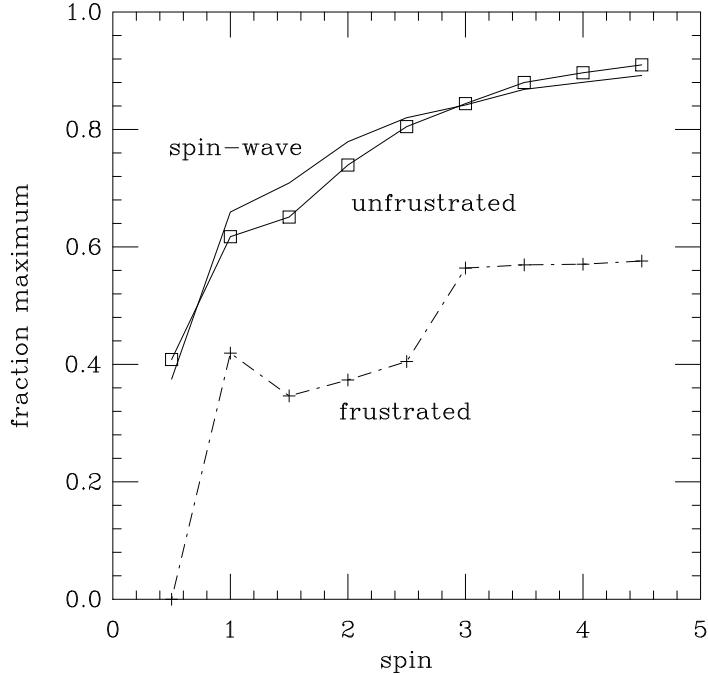
{\bf Figure \figchis.}  Exact diagonalization and spin-wave theory for
$J$=$1$.
The middle trace, marked with
squares and labeled ``unfrustrated,'' plots $\sqrt{\chi(0)\chi(2)}$,
normalized to its maximum possible value,
on the eight-site cluster as a function of spin (2 is as far away as
two plaquettes get on this cluster; results for $\sqrt{2}$ do not
differ markedly).  Above it (solid line), we plot $\chi$ from
spin-wave theory.  The close agreement of the two bolsters our
confidence in the results to some extent.  Below these two lines,
we plot $\sqrt{\chi(0)\chi(\sqrt{2})}$ on the {\it frustrated\/} eight-site
cluster.
}

For spin 1, we found in the 16-site cluster no ferromagnetic transition
for any value of $J$ up through $100$.
Instead of dropping suddenly,
the chirality-chirality correlation in figure \figchij\ 
falls continuously to zero
as $J$ approaches about 6.
While we cannot, as for spin \frac1/2, write the quartic term as purely
ferromagnetic, it can be written simply as the sum of a constant, a
ferromagnetic piece, and a site-interchange operator,
($1 - \bold S_i\cdot \bold S_j + \roman X_{ij}$).\fft{ %
Alternatively, one can
write the quartic interaction between quantum spin-1
coherent states (ground-state eigenvectors
of operators $-\bold S\cdot{{\bf\hat n}(\theta, \phi)}$ for some 
$\hat{\bold n}$ on the unit sphere) as the sum of again a
ferromagnetic piece, a constant,
and a Heisenberg-squared term favoring chirality:
$\langle(\bold S_0\cdot\bold S_1)^2\rangle=
-(1/2)\cos\psi + 5/4 + (1/4)\cos^2\psi$, with $\psi$ the relative
angle between the two coherent states.
\ftname(cohere)
} %
We interpret the falling order as a ferromagnetic tendency
that still carries some weight at spin 1 but vanishes in the limit of
infinite spin.

Figure \figchis\ displays
the spin dependence of the largest-distance chirality-correlation
function achievable on the unfrustrated eight-site geometry for spins between
\frac1/2 and \frac9/2; all correlations are shown normalized to the
largest quantum-mechanically possible value for $\chi$ on three
sites, and $J$ is fixed at $1$.
For comparison, we also show the results of the spin-wave
calculation for the correlation at infinite separation, while contrasting
the normalized chirality-chirality correlation for a frustrated geometry
in which even the classical system cannot have full chiral order on every
plaquette.

These results cannot be considered conclusive, given the small cluster
sizes and the inability (with so few sizes from which to choose) to 
apply finite-size scaling.  However, they at least support the suspicion
that the system might have non-trivial long-range order\ft{%
That is, larger than the zero-temperature ordering of uncorrelated
sublattices, in a regime of $J$'s for which the trivial mechanism
should fail to operate.
} %
for spins greater than \frac1/2.

\header{Spin-wave calculation}

Spin-wave theory (for a review, see Manousakis, 1991) assumes
a ground state close to the
classical state, with dynamical variables, supposed small, indicating
deviations.  The theory constitutes an expansion in these small,
dimensionless variables divided by the spin, which we wish to vary.  As
such, it cannot say anything about the properties of a system
in a regime where the assumed ground state fails to approximate
the true state, but it does give a
rough idea of where the failure occurs, in our case the approximate
spin value below
which the classical chiral ground state is no longer an
appropriate starting point.

We diagonalize not in the traditional Holstein-Primakov Bose operators
but in a more convenient linear combination of them.  Our methods,
while not differing in spirit from usual treatments, may be of use to
others, so we outline them in the appendix.

As figure \figchis\ shows, the spin-wave results agree fairly closely
with the finite-size numerical correlations.  The two calculations measure
somewhat different objects.  The diagonalization looked at
$\langle\chi(0)\chi(r)\rangle$ for as large an $r$ as could fit in
a cluster.  It could not directly measure the ``chiral polarization''
$\langle\chi(0)\rangle$,
since there is nothing to break the symmetry
of positive and negative chirality.  The spin-wave calculation, in
contrast, breaks that symmetry from the outset and asks how large
the deviation is: it does measure\fft{%
\emergencystretch1cm%
The chiral-chiral correlation in the spin-wave
theory for this problem, $\langle\chi(0)\chi(r)\rangle$, is essentially the
square of this, approximately independent of $r$.  In other words,
the deviation-deviation correlation is negligible.  This is confirmed
by looking at the deviation $\Delta\chi(\bold k)$ as a function of wavenumber
$\bold k$: it is nearly flat, giving a deviation correlation sharply
peaked at $\bold 0$ in real space.} %
$\langle\chi(0)\rangle$.
A deviation from order calculated by the spin waves greater than
the maximum value of the order itself would indicate the
breakdown of spin-wave theory and probably the loss of long-range order.
The figure shows the chirality for $J=1$ in all
cases greater than zero, although down at spin \frac1/2 the deviation
from full chirality is more than half.

The actual value $s$ of spin enters into spin-wave theory only
in determining when the theory breaks down ({\it viz.}, when the
deviations grow larger than the maximum $\chi$);
the system
can't know about the quantum-mechanical
ferromagnetic tendency of $(\bold S_i\cdot\bold S_j)^2$
at small spins.  Thus we would expect the spin-wave calculation to
underestimate the deviations for $s=1/2$.

\header{Is the order long-ranged?}

The spin-wave theory of figure \figchis\ fails to find any value of
spin for which fluctuations completely destroy chiral order, although
at spin \frac1/2 {\it most\/} of the order has been eroded.
Considering the uncertainties (see the appendix) of extending spin-wave
theory to spin \frac1/2, and the lack of order at $J\geo1$
in the diagonalizations, we may well believe
that the spin-\frac1/2 system is {\it not\/} ordered in this regime.

For spin 1, it is easier
to come to a conclusion.  While 
figures \figfit c and \figfit a suggest a greater extrapolated order for spin
1 than for spin \frac1/2,
as well as a longer decay length, the statistics do not
permit a clear interpretation.
However, the unambiguously ordered prediction of
spin-wave theory, combined with reasonably close quantitative agreement
of the two calculations, suggests that spin 1 is ordered.  Since
the peak ordering occurs at $J=1$
we can be confident the effect is not caused by fluctuations of
uncoupled lattices, as with spin \frac1/2.
(We shall estimate that the lattices should couple for $J\geo0.5$ or even
for smaller $J$.)
We believe that higher spins are ordered as well.

The Ising-like character of the chiral order parameter further bolsters
the conclusion.  Except in the identified ferromagnetic state found
for large enough $J$ when spin is \frac1/2, we see in the
exact diagonalization antiferromagnetic
order on the antiferromagnetically-coupled sublattices.  Since
we expect Ising-like order to be at least as robust as spin order,
we would have to be surprised by the absence of the former in the
presence of the latter.
Only when spin is \frac1/2 do we find a
transition to a competing magnetic state, which then destroys
chirality.

\header{No non-trivial long-range order for spin-\frac{\bf1}/{\bf2}}

We consider another approach to the question of spin \frac1/2
in the troublesome regime of smaller $J$, where lattice decoupling
appeared to induce the ``trivial'' ordering of footnote \sqa.
(For larger $J$, a ferromagnetic transition killed even that kind of
ordering.)  We will present evidence that any infinitesimal temperature
should destroy chiral ordering for spin \frac1/2.  This is a stronger
statement than the analogous assertion for magnetic ordering in the
Heisenberg model for two reasons: chirality has an Ising-like rather
than a continuous symmetry, so the Mermin-Wagner theorem does not apply;
furthermore, we shall find that infinitesimal temperature kills
even the local chirality on individual plaquettes, not just chirality
correlation.

Restricted to spins \frac1/2, the Hamiltonian becomes (up to a constant)
$$
H= 
4\sum_{\eightpoint\hbox to2em{\hss${\hbox{anti-ferro.\ }%
^{\vphantom{2}}i,j}$\hss}}
\bold S_i\cdot\bold S_j
- 8 J \sum_{\eightpoint\hbox to2em{\hss${\hbox{Heisenberg}^2\ i,j}$\hss}}
\bold S_i\cdot\bold S_j\smash{\quad,}
\eq(frust)
$$
the {\it classically\/} frustrated
behavior of which we may scrutinize for hints about the quantum system.
One cannot always hope to guess the correct ground state for a frustrated
classical system, so we used a straightforward numerical minimization procedure
on finite clusters of classical spins with periodic boundary conditions
(open boundaries produced spurious chiral phases).

Equation \(frust) follows from \(Hamiltonian) by an operator identity,
not by an approximation.  The approximation consists in treating the
spins in \(frust) classically.\fft{ %
\emergencystretch3em%
This treatment is identical
to a variational calculation in the subspace of
spin-\frac1/2 co\-herent-state products (see footnote
\cohere).
}
Letting the classical spins have unit length (instead of length $\frac1/2$),
we divide the coefficients in \(frust) by four.

Numerically minimizing the energy of the configuration with respect to the
angles of the classical spins, we find three regimes of $J$.
Without ferromagnetic bonds ($J=0$), the system consists of
three antiferromagnetic lattices with undetermined relative
angles, a degeneracy that persists up through
a $J$ of approximately 0.5, when states with mixed chiral order
(possibly long ranged) take over: these states differ from the
ferromagnetic chirality we saw at higher spins in alternating
positive, negative, and zero values on adjacent plaquettes.
Similarly,
the ferromagnetic state that prevails at infinite $J$
(turning off the antiferromagnetic bonds) appears to be stable against
canting deformations (at zero temperature) down to a $J$ of around 0.97,
in remarkable agreement with the diagonalizations of figure \figchij.

These classical results suggest no chiral order of the
sort we seek, but we can go a little further by examining the effects
of quantum fluctuations on the undetermined ground state we found for
small $J$.
Before addressing quantum
fluctuations directly, however, we will look at their classical
relatives, thermal fluctuations.

\header{\it low-temperature expansion}
In the regime $0<J\leo0.5$, where the classical model found a degenerate
ground state,
we wish to examine whether fluctuations will favor one
relative sublattice orientation over the others.
The classical theory of multiple phase transitions begins with
a low-temperature
expansion (Domb, 1960, Fisher and Selke, 1981), which will
serve as an inspiration for our rather simpler
purposes.
Non-zero temperature will induce low-energy excitations, which we may
view either as long-wavelength twists in the antiferromagnetic lattices
or as very small deviations, tips, of isolated spins (in an Ising
model, these would be ``flips'' costing a discrete energy; see
Fisher and Selke, 1981).  Because of
the analogy to the Ising model, we present
the calculation from the second viewpoint, although we have arrived at
the same result the first way, too.

A single isolated tipped spin will not prefer any relative lattice
orientation, since each ferromagnetic bond linking the affected
spin with a spin in another sublattice is canceled by an equal
bond to a third spin, pointing exactly oppositely, on that other sublattice.
When two neighboring spins on different sublattices tip simultaneously,
however, we shall see that they favor a particular orientation of their
two sublattices.

Consider tips of a single spin on the even sublattice ($0$) (refer
to figure \figspace) and of a
neighboring spin on either of the others, say $1$.
(One obviously gets the same results for sublattices $0$ and
$2$ and in fact also for sublattices $1$ and $2$.)
We fix the untipped spins in sublattice $0$ to point in
the positive and negative ${\bf\hat z}$
directions, while sublattice $1$ points some angle $\theta_1$
from the ${\bf\hat z}$ axis
toward the ${\pm\bf\hat x}$ direction.  See figure \figentropic.  If
the tipping angles away
from their respective preferred axes are both some small number $\delta$,
while the azimuthal angles {\it relative to the sublattice directions\/}
are $\phi^{(0)}$ and $\phi^{(1)}$, we
can add up all the broken bonds to get a cost for the excitation of
$$
\Delta E =
4 \delta^2
+
2 J \delta^2 ( \cos\theta_1 \cos\phi^{(0)}\cos\phi^{(1)}
+ \sin\phi^{(0)} \sin\phi^{(1)} )
\smash\quad,
\eq(spintip)
$$
where we have kept only the leading power of the small tipping
angle $\delta$.

\efigureinsert{figentropic.ips}{\capentropic}{\epsfysize=3in}{%
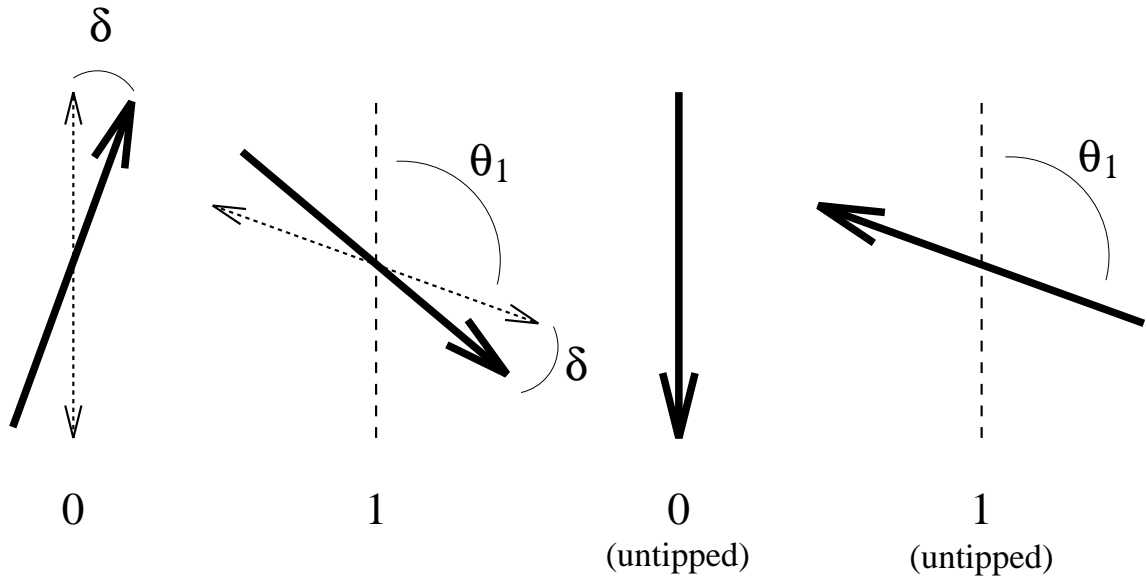
{\bf Figure \figentropic.}  Low-temperature expansion: we consider
two of the three sublattices.  Spins on sublattice $0$ point straight
up or down (one is shown as ``untipped''), while those on sublattice
$1$ point an angle $\theta_1$ away from the vertical.  Thermal fluctuations
sometimes cause two neighboring spins to tip an angle $\delta$ away from
their equilibrium orientations, indicated for the two spins on the left
by double-arrowed thin lines.  Not pictured are the azimuthal angles
$\phi^{(0)}$ and $\phi^{(1)}$ about these thin lines.
}

The first term represents the cost of one spin tip on each of the two
otherwise antiferromagnetic lattices;
it will serve as an expansion parameter
if we can fix $\delta$
at some small
value.
(Without this approximation, connected tips of arbitrarily many
spins by arbitrarily small $\delta$'s would carry the same weights,
something we cannot handle.
There is some subtlety in the
order in which temperature,
$\delta$, and the quantization of
azimuthal angles go to zero as well
as in the linked-cluster theorem; see
Domb, 1960.)

Given an excitation cost, we add Boltzmann weights from all
values of the
angles $\phi^{(0)}$ and $\phi^{(1)}$
to get a change
in the free energy, due to the possibility of such excitations, proportional
to
$$
\eqalign{
f(\theta_1) &= -\kbt
	\int_{-\pi}^{\pi} d \phi^{(0)} \int_{-\pi}^{\pi} d\phi^{(1)}
	\exp(-\beta\Delta E(\phi^{(0)},\phi^{(1)})) \cr
&= -4\pi^2 \kbt
	e^{-4\beta\delta^2} I_0(\alpha) I_0({b\over4\alpha})\smash\quad,\cr
}
\eq(free)
$$
\vskip0.25\baselineskip\noindent%
where $\alpha=\sqrt{(a + b/2 + \gamma \cos\theta_1)/2}$,
$a=\gamma\cos^2\theta_1$, $b=\gamma\sin^2\theta_1$,
$\gamma=4\beta^2 J^2 \delta^4$, and $I_0$ is a Bessel function of imaginary
argument.
One verifies numerically or graphically
that $f(\theta_1)$ has its minima at $0$ and $\pi$, meaning that
sublattices $0$ and $1$ prefer, in any infinitesimal temperature, to
be aligned.\ft{%
\emergencystretch2cm\tolerance100000%
One may numerically integrate out the parameter $\delta$ without changing
this con\-clu\-sion.} %
Since the same is true of $0$ and $2$ and of $1$ and $2$,
every spin in the problem will point very nearly in plus or minus
the same direction, and there will be {\it no chirality}, not even
locally.

\header{\it treating quantum fluctuations as perturbations}
We now replace thermal fluctuations with those due to quantum mechanics, for
which purpose
we expand effectively in lattice fluctuations.  For antiferromagnetic
fluctuations, one typically
writes
$\bold S_i \cdot \bold S_j = S_i^z S_j^z 
+ {1\over2}\lambda ( S_i^+ S_j^- + S_i^- S_j^+ )$,
where quantum mechanics requires that $\lambda$ should equal unity, while
$\lambda=0$ is essentially classical.

We shall later see that fluctuations away from the N\'eel state, present
in the antiferromagnetic sublattices, will order the randomly-oriented
lattices.  First, however, we consider fluctuations in the ferromagnetic
(spin-\frac1/2-Heisenberg-squared) bonds.

Isolated fluctuating spins in a given sublattice cannot break the
degeneracy, but as in the thermal case, connected fluctuations of
neighboring spins on different sublattices can.  Consider sublattice
$0$ to be quantized in the $\bf\hat z$ direction, while the spins in
sublattice $1$ are quantized along an axis an angle $\theta_1$ off
$\bf\hat z$.  The spins in this second sublattice are in pure states
only if we rotate by $\theta_1$ the coordinates we use to describe them:
thus for site $i$ in sublattice $0$ and site $j$ in sublattice $1$,
$$
\eqalign{
&\bold S_{0i}\cdot\bold S_{1j} =\cr
&\quad \left\{\cos\theta_1 S_{0i}^z S_{1j}^{'z}\right\}\cr
&\quad+{\lambda\over2}\left\{-\sin\theta_1 S_{0i}^z [ S_{1j}^{'+} + S_{1j}^{'-} ]
	+\sin\theta_1 [ S_{0i}^+ + S_{0i}^- ] S_{1j}^{'z} \right\}\cr
&\quad+{\lambda\over4}\left\{
(\cos\theta_1-1) [ S_{0i}^+ S_{1j}^{'+} + S_{0i}^- S_{1j}^{'-} ]
+(\cos\theta_1+1) [ S_{0i}^+ S_{1j}^{'-} + S_{0i}^- S_{1j}^{'+} ] \right\}
\smash\quad,\cr
}
\eq(rotated)
$$
where the primes indicate rotated operators.  The first term is just
the classical Ising piece.  We have multiplied the remaining
``quantum-mechanical'' terms by $\lambda$.\ft{%
Setting $\lambda=1$ eliminates the anisotropy essential to this
calculation, but as the parameter is tuned closer to its true
quantum mechanical value, one would also have to turn on the antiferromagnetic
fluctuations.
} %
Each of the second set of terms flips
a single spin, thereby contributing to the energy shift
in second-order Rayleigh-Schr\"odinger
perturbation theory.
However, each spin acted on by one of the raising or lowering operators
in these terms has one unperturbed neighbor (acted
on by $S_{0i}^z$ or $S_{1j}^{'z}$)
pointing up in the appropriate lattice and one neighbor pointing down (here
we ignore antiferromagnetic fluctuations).
These sum to zero; therefore none of the terms in the second set contributes.

Each of the terms in the third set of braces contributes (with the same
energy denominator) in second-order perturbation theory; the energy
shift is proportional to
$$
- (\cos\theta_1 - 1 )^2 - (\cos\theta_1 + 1 )^2
~\propto~ -\cos^2\theta_1 - \hbox{const}\smash\quad,
\eq(perturb)
$$
which lowers the energy most when $\theta_1=0,\ \pi$.
Quantum fluctuations, therefore, like the classical thermal ones, predict
alignment of the sublattices and hence a lack of chiral ordering in the
regime considered.

\header{\it A mean-field theory agrees, too}
In a similar spirit, consider the four spins shown in figure \figfoursite:
spins $1$ and $2$ sit on one sublattice and are joined by an antiferromagnetic
bond, while $3$ and $4$ sit on another sublattice and are similarly joined.
Ferromagnetic bonds link spins across sublattices, while a mean field
$ \pm 12 h \bf\hat z $ keeps $1$ and $2$ pointing in the $\pm\bf\hat z$
directions, while a rotated version of the same keeps $3$ and $4$ pointing
in the $\pm(\cos\theta_1{\bf\hat z} + \sin\theta_1{\bf\hat x})$ directions.
(The $12$ comes from three antiferromagnetic bonds per site not included
in figure \figfoursite\ but contributing through the mean field).
We assume the mean {\it ferromagnetic\/} field is small enough to be ignored,
both because $J$ is small and because the contributions from neighbors
will nearly cancel.  Treated classically, this cluster shows no preference
for any relative orientation of the sublattices.  Quantum-mechanical
fluctuations, however, once again break the degeneracy.

We (or rather our computers) easily diagonalize the four-spin Hamiltonian with
the extra mean-field terms; we then set $h$ equal to the
staggered magnetization and iterate until $h$ converges.  For $J$ smaller
than about $1.6$ (quite a large value), $h$ converges to a non-zero
value not too far from full antiferromagnetic sublattice ordering.
Plotting the total energy as a function of $\theta_1$, we again find
a roughly sinusoidal dependence, with the minimum at $0$ (equivalently $\pi$),
confirming that the lattices prefer alignment over chirality.\ft{%
Insofar as low-order perturbation theory on an infinite lattice may be 
viewed as a calculation on a small cluster (Gelfand {\it et al.}, 1990),
this mean-field computation, examining antiferromagnetic fluctuations,
complements the previous perturbative expansion
in ferromagnetic fluctuations.} %

\efigureinsert{figfoursite.ips}{\capfoursite}{\epsfysize=3in}{%
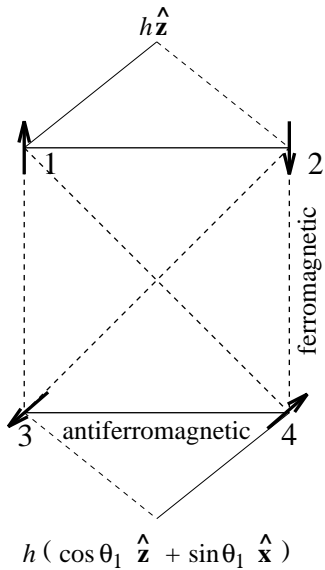
{\bf Figure \figfoursite.}  Quantum mean-field theory: consider
spins $1$ and $2$ on one sublattice and $3$ and $4$ on another.  The
two spins on each sublattice are linked by an antiferromagnetic bond
(solid lines), while spins on opposite lattices are linked ferromagnetically
(weakly).  To account for antiferromagnetic bonds to spins external to
this diagram, we apply mean fields proportional to $h\bf\hat z$ and
$h(\cos\theta_1{\bf\hat z}+\sin\theta_1{\bf\hat x})$ on each sublattice,
determining the self-consistent value of $h$ iteratively.
}

This mean-field spin-\frac1/2 result, along with the preceding
approximations finding no order for
small $J$, suggests that the low peaks of the cluster diagonalizations
(figure \figchij) represent no chiral correlations beyond the random
``background'' (footnote \sqa).

\header{\it spin $>\hbox{\it1}$ order}
We easily repeat the mean-field calculation for all spins 1 through \frac5/2 on
the cluster
of figure \figfoursite\ with the full Hamiltonian, equation \(Hamiltonian),
and find, in contrast to the spin-\frac1/2 case, that
pairs of sublattices prefer to be mutually {\it perpendicular}.  (The
minimum energy occurs at a relative orientation of $\pm\pi/2$.)  This
corroborates the conclusions based on the larger diagonalizations and
on spin-wave theory, {\it viz.}\ that spins 1 and higher likely
support long-ranged non-trivial chiral
order in our model, while spin-\frac1/2 does not.

If so simple a model as the present, concocted to make chiral ordering
easy, fails to show that order for spins-\frac1/2, we would view cautiously
extrapolations of chirality to spin-\frac1/2 electrons in other square-lattice
models.  On the other hand, spin-chiral order would not surprise us when
electrons combine into spin-1 or higher objects.

\bigskip
\lheader{Acknowledgements}
This work was supported in part by a Department of Energy Distinguished
Postdoctoral Fellowship.

\header{Appendix: linear spin-wave theory}
\header{\it choice of expansion parameters}
Classically there are infinitely many ground states, each with
a unit cell of eight sites, all related by
simple spin rotations or reflections.  Choosing one of these,
we construct a local right-handed coordinate system
on each site in 
which $S_i^z$ ($i$ marking both the site, 0 through 7, within a unit cell
and that cell's coordinate) gives the component of
spin in the classically-favored direction, with $q_i\equiv S_i^x/s^{1/2}$
and $p_i\equiv S_i^y/s^{1/2}$
presumably relatively small deviations ($s$ is the spin).
We
choose the names $q$ and
$p$ because in fact these deviations are approximately conjugate:
$$
[q,p] ~=~ %
i S_i^z / s
~=~ i \hbox{\rm~~minus~~first-order small corrections}.
\eq(qmconj)
$$
From this approximate correspondence follow the semiclassical (linear
spin-wave) equations of motion,
$$
\eqalign{\dot q_i &\approx \hphantom{-}{{\partial H}\over { \partial p_i } }\cr
	 \noalign{\vskip0.25\baselineskip}
         \dot p_i &\approx -{ {\partial H } \over { \partial q_i } }
	 \smash{\quad.}\cr}
\eq(motion)
$$
(Alternatively, consider each spin $\bold S$ as a %
symmetric classical top, fixed at the base and with center
of mass at $\bold R$, proportional to $\bold S$.  If the
mass is vanishingly small [and the angular velocity infinite,
so as to yield angular momentum $\bold S$], we can ignore
nutation.  Taking the
torques resulting from all the other spins in the system to act
at $\bold R$, we have
$\dot\bold S = - \bold R\times\partial H/\partial\bold R
 = -\bold S\times\partial H/\partial\bold S$, from which
we get \(motion) with exact equality, within this
classical approximation.)

Armed with first-order expressions for the $x$ and $y$ components of
spin, we can define the linear approximation to the $z$ component as
having the form
$$
S_i^z = s' - {1\over2s''}({S_i^x}^2 + {S_i^y}^2)\smash{\quad.}
\eq(sz)
$$
While it would be tempting to fix the constants $s$ and $s'$ by the
requirement
$$
S^2 ~=~ {S^x}^2 + {S^y}^2 + {S^z}^2 ~=~ s(s+1)\smash{\quad,}
\eq(s2)
$$
the resulting substitution disagrees with conventional spin-wave theory at
small spin (they agree of course in the limit of large $s$).  The
discrepancy apparently arises in the choice of which constraints to
impose exactly and which only to leading order in the perturbations.
Holstein-Primakov satisfies all the
important constraints
simultaneously; we therefore
relate our $q$ and $p$ to the operators of conventional spin-wave
theory.

The usual Holstein-Primakov spin-wave treatment
identifies the spin operators $S^-$
and $S^+$ approximately
with Bose creation and annihilation operators $a^\dagger$
and $a$, making the deviation from full magnetization ($S^z=s$) look
like a Bose excitation, called a magnon.
So long as the magnon-number operator $a^\dagger a$ is small
compared to the $S^z$ quantum number, the approximation is good.  Of course,
the Bose excitation spectrum is unbounded above, while enough applications
of $S^-$ to any state will annihilate it.  There are therefore corrections
to the linear theory of quadratic order in the magnon number.  Specifically
(see Manousakis, 1991),
$$
S_i^+ = \sqrt{2s}(1 - a_i^\dagger a^{\phantom{\dagger}}_i)^{1/2}
a^{\phantom{\dagger}}_i
 = \sqrt{2s}(1 - {1\over2}a_i^\dagger a^{\phantom{\dagger}}_i
+ \dots.) a^{\phantom{\dagger}}_i\smash{\quad.}
\eq(HP)
$$
Inserting the resulting expressions
for $S^x$ and $S^y$ into the form \(sz) (discarding pieces of order
$(a^\dagger a)^2/s$) results in the identification
$$
s' = s'' = s + 1/2\smash{\quad;}
\eq(method4)
$$
this choice guarantees that our linear spin-wave theory reproduces the
usual results for the antiferromagnet.
(We note that Mattis, 1988, equation 5.9, assigns $s'=s+1/2$ but
$s''=s$.  In contrast, equation \(s2) implies the assignment
$s'=s''=\sqrt{s(s+1)}$,
which however poorly approximates the known magnetization in the
$s\rightarrow1/2$ extrapolation.)

We use $q$ and $p$ instead of $a^\dagger$ and $a$ because of the
relative simplicity of expansions of the quartic terms in the
Hamiltonian \(Hamiltonian).

\header{\it solving the classical problem}
We truncate the Hamiltonian at terms quadratic in the operators
$p_i$ and $q_i$; equation \(motion) then describes a classical,
harmonic problem.  Consider all the $p$'s and $q$'s in a single
unit cell (say $\bold r$), putting them together in the vector
$$
{\bfit\sigma}^{\roman T}_{\bold r}
= ( q_{\bold r,0}\ q_{\bold r,1}\ \dots
\ q_{\bold r,7}\ 
p_{\bold r,0}\ \dots\ p_{\bold r,7} )\smash{\quad,}
\eq(sigma)
$$
where $\roman T$ indicates transposition, and the notation in the subscripts
now separates the cell position and the index (0--7) within the unit cell.
We assume a spin-wave mode $\bold k$ with
$$
\sigma_{\bold r,i}
= \hbox{\rm Real}\,(
\sigma_i(\bold k) \exp(i \bold k\cdot\bold r - i \omega(\bold k) t)
)
\eq(sigmak)
$$
($t$ is time), giving the eigenvalue equation
$$
\omega(\bold k){\bfit\sigma}(\bold k)
= \roman T(\bold k){\bfit\sigma}(\bold k)\smash{\quad,}
\eq(eigen)
$$
where the non-Hermitian matrix
$\roman T$ can be determined directly from
\(motion) and \(sigmak) but in fact is also
related to
the quadratic form $\roman M$ of the Hamiltonian,
$$
H = \hbox{\rm constant} ~+~ \sum_{\bold k}
{\bfit\sigma}^\dagger(\bold k)\roman M(\bold k){\bfit\sigma}(\bold k)
\smash{\quad,}
\eq(M)
$$
by
$$
\roman T(\bold k) = 2 i \roman A \roman M(\bold k)\smash{\quad;}
\eq(T)
$$
here $\roman A$ is the $16\times16$ matrix (see Goldstein, p.347)
$$
\roman A = \left(%
\vcenter{%
\halign{%
\hfil#\ &\hfil\ #\ \cr%
$\bf\hphantom{-}0$ & $\bf\hphantom{-}1$ \cr%
$\bf-1$            & $\bf\hphantom{-}0$ \cr%
}%
}%
\right )\smash{\quad.}
\eq(A)
$$

Once we know the eigenvectors of $\roman T$,\fft{%
It speeds the numerical calculation to rewrite \(eigen) as
$2 i \roman M^{1/2} \roman A \roman M^{1/2} \bold y = \omega\bold y$, where
${\bfit\sigma} = \roman M^{-1/2}\bold y$ and $\roman M^{1/2}$ is chosen
to be a Hermitian square root of the Hermitian matrix $\roman M$ (which
we can do when, as for most of this work, $\roman M$ is positive
definite).  This new eigenvector equation in $\bold y$ is then Hermitian.
} %
we can normalize them so that each
contains energy $\omega/4$ (since every independent mode corresponds
to two time-reversed eigenvectors).
Then the substitution \(sigmak)
gives directly any observable expressed in terms of $q$'s and $p$'s.
For instance, remembering to use \(method4) and for a particular
choice of coordinate system and labeling of points within the
unit cell, one finds on the plaquette whose vertices are $0$,
$3$, and $1$,
$$
\chi_{\scriptscriptstyle031} = 
s^3 + {3\over2}s^2 
- {1\over2}s^2\left\{%
\vert q_0+p_3\vert^2+\vert p_0-p_1\vert^2+\vert q_1+q_3\vert^2
\right\}\smash\quad.%
\eq(chi031)
$$

\header{\it alternative solution employing a canonical transformation}
An equivalent approach more transparently connected to the standard
Holstein-Primakov treatment (or to that
of White {\it et al.}, 1965, but applied directly to the
variables $p$ and $q$) begins with the Fourier transform, \(M).
Then $\dot{\bfit\sigma} = \roman A\partial H/\partial{{\bfit\sigma}^\dagger}
= 2 \roman A\roman M{\bfit\sigma}$, from which\fft{%
The 2 comes from the interdependence of ${\bfit\sigma}$
and ${\bfit\sigma^\dagger}$.} %
\(eigen) again follows. 
We form the columns of the matrix $\roman V$ from the eigenvectors
of $\roman T$; $\roman V^{-1}\roman T\roman V$ is diagonal,
implying by the equations of motion that $\roman V^\dagger\roman M\roman V$
is also diagonal.  
A few lines of algebra demonstrate furthermore
that $\roman V^\dagger\roman A\roman V$
is diagonal.\fft{%
We assume all the normal-mode frequencies are distinct.  Let $\bold v_i$ be an
eigenvector of $\roman T$.  Then
$\roman M\bold v_i = (1/2) i \omega_i \roman A \bold v_i$, while
$\bold v_i^{\dagger}\roman M = (1/2) i \omega_i \bold v_i^{\dagger} \roman A$
(since frequencies are real and $\roman M$ Hermitian).  For two different
eigenvectors,
$\bold v_i^{\dagger} \roman M \bold v_j
= (1/2) i\omega_i \bold v_i^{\dagger} \roman A \bold v_j$
but also equals
$ (1/2) i\omega_j \bold v_i^{\dagger} \roman A \bold v_j$, implying,
since $\omega_i\not=\omega_j$,
that $\roman V$ diagonalizes $\roman A$ (and $\roman M$ as well).
} %
We can without changing any of the foregoing normalize
and permute the columns of $\roman V$ so that
$$
\vcenter{%
\halign{%
\hfil#\hfil&%
\hfil$\displaystyle\relax#$&%
$\displaystyle~=~\hbox{\rm diag}($\hfil$\displaystyle\,#\,$\hfil&%
\hfil$\displaystyle\,#\,$\hfil&%
$\displaystyle\dots$\hfil$\displaystyle\,#\,$\hfil&%
\hfil$\displaystyle\,#\,$\hfil&%
$\displaystyle\dots$\hfil$\displaystyle\,#\,$\hfil%
$\displaystyle)$&#\cr%
\empty&\roman V^{-1}\roman T\roman V
&\omega_1&\omega_2&\omega_8&-\omega_1&-\omega_8\cr
and&\roman V^\dagger\roman A\roman V
&-i&-i&-i&i&i&\smash{\quad,}\cr
}%
}%
\eq(notsymplectic)
$$
a step that corresponds to finding a set of uncoupled harmonic-oscillator
variables with the masses all set to \frac1/2 and the spring constants
to $2$.
Now \(notsymplectic) is not the
condition required of a canonical transformation, but
$\roman V'=\roman V\roman W$, where
$$
\roman W = {1\over\sqrt{2}} \left(
\vcenter{%
\halign{%
\hfil#\ &\hfil\ #\ \cr%
$\bf\hphantom{-}1$ & $\bf\hphantom{-}i$ \cr%
$\bf\hphantom{-}1$ & $\bf-i$            \cr%
}%
}
\right)
\eq(Wd)
$$
satisfies the symplectic condition,
$$
\roman V^{\prime\dagger}\roman A\roman V' = \roman A\smash{\quad,}
\eq(symplectic)
$$
while still diagonalizing $\roman M$.
The transformation
$$
{\bfit\sigma} = \roman V'{\bfit\sigma'}
\eq(canonical)
$$
is therefore canonical, meaning that it preserves
all the commutators of $p$'s and $q$'s.  The fact that $\roman V'$ also
diagonalizes $\roman M$ lets us write the Hamiltonian $H$ as
the sum of uncoupled oscillators, which we know how to solve.

We now rewrite the original $q$'s and $p$'s in terms of the
Holstein-Primakov operators $a$ and $a^\dagger$ ({\it e.g.},
$q_i(\bold k) = (1/\sqrt{2})(a_i^\dagger(-\bold k)
+ a_i^{\vphantom{\dagger}}(\bold k))$,
and then derive a simple expression in terms of the
matrix $\roman V'$ for the ground-state expectation value
$(1/2)\langle\sigma_i^*\sigma_j^{\vphantom{*}}\rangle$, which we use
to evaluate expressions such as \(chi031).  The results of course
are identical to those we derived first.

\vfill\eject
\header{References}
{
\frenchspacing
\leftskip=\the\parindent
\multiply\parindent -1
\def\J#1{{\it #1}}\nobreak%
\def\BOOK#1{{\it #1}}\nobreak%

Affleck, I., T. Kennedy, E.H. Lieb, and H. Tasaki, 1987,
\J{Phys. Rev. Lett. \bf59}, 799.

Agterberg, D.F. and M. B. Walker, 1994,
\J{Phys. Rev. \bf B50}, 563.%

Anderson, P.W., 1963,
in F. Seitz and D. Turnbull,
eds., \BOOK{Solid State Physics \bf14}, 99.

Baskaran, G., 1989,
\J{Phys. Rev. Lett. \bf63}, 2524.

Chang, K., I. Affleck, G.W. Hayden, and Z.G. Soos, 1989,
\J{J. Phys.:Condens. Matter \bf1}, 153.%

Domb, C., 1960, \J{Adv. Phys. \bf9}, 149.

Fisher, M.E. and W. Selke, 1981, \J{Phil. Trans. Roy. Soc. (London) \bf302}, 1.

Gelfand, M.P, R.R.P. Singh, D.A. Huse, \J{J. Stat. Phys. \bf59}, 1093.

Goldstein, H., 1981, \BOOK{Classical Mechanics}, 2nd. ed., Addison-Wesley,
Reading, Mass.

Gooding, R.J., 1990, ``Exact mapping of the $J_1, J_2$ Frustrated
Square QAFM into a Chiral Spin Liquid Hamiltonian,'' \J{unpublished}.

Gor'kov, L.P. and A. Sokol, 1992,
\J{Phys. Rev. Lett. \bf69}, 2586.%

Harris, E.A. and J. Owen, 1963,
\J{Phys. Rev. Lett. \bf11}, 9.

Henley, C.L., 1989, \J{Phys. Rev. Lett. \bf62}, 2056.

Korshunov, S.E., 1993,
\J{Phys. Rev. B\bf47}, 6165.

Manousakis, E., 1991,
\J{Rev. Mod. Phys. \bf 63}, 1.%

Marston, J.B. and I. Affleck, 1989,
\J{Phys. Rev. B\bf39}, 11538.

Marston, J.B. and C. Zeng, 1991, \J{J. Appl. Phys. \bf69}, 5962.

Masui, S. and D.D. Betts, 1993,
\J{Phys. Rev. B\bf48}, 6225.

Mattis, D.C., 1988, \BOOK{The Theory of Magnetism I}, corrected 2nd printing,
Springer, Berlin.

Mitra, S.N. and K.G. Chakraborty, 1994,
\J{J. Phys: Condens. Matter \bf6},
10533.

Ramasesha, S. and Z.G. Soos, 1984,
\J{Intl. J. Quant. Chem. \bf 25}, 1003--1021.

Takhtajan, L.A., 1982, \J{Phys. Lett. \bf87A}, 479.

Villain, J., 1979, \J{Z. Phys. B\bf33}, 31.

Villain, J., R. Bidaux, J.P. Carton, R. Conte, 1980,
\J{J. Phys. (Paris)\bf41}, 1263.

Wen, X.-G., F. Wilczek, and A. Zee, 1989, \J{Phys. Rev. B\bf39}, 11413.

White, R.M., M. Sparks, I. Ortenburger, 1965,
\J{Phys. Rev. \bf139}, A450.

}

\vfil\end